\documentclass[aps,amsmath,amssymb,prl,showpacs,twocolumn]{revtex4}

\usepackage{graphicx}    
\usepackage{color}
\usepackage{bm}

\newcommand{\be}{\begin{equation}}
\newcommand{\ee}{\end{equation}}
\newcommand{\ba}{\begin{eqnarray}}
\newcommand{\ea}{\end{eqnarray}}
\newcommand{\re}{\mathrm{e}}            

\newcommand{\Ca}{\mathrm{Ca}}

\begin{document}

\title{Step-emulsification in nanofluidic device}
\author{Z. Li$^1$}
\author {A.\ M.\ Leshansky$^{2}$}
\email{lisha@technion.ac.il}
\author{L.\ M.\ Pismen$^{2,3}$}
\author{P. Tabeling$^1$}
\affiliation{$^1$MMN, CNRS, ESPCI Paris-Tech, 10 rue Vauquelin, 75005 Paris, France \\
$^\mathrm{2}$Department of Chemical Engineering and Russell Berrie Nanotechnology Institute, Technion -- IIT, Haifa, 32000, Israel\\
$^\mathrm{3}$Minerva Center for Nonlinear Physics of Complex Systems, Technion -- IIT, Haifa 32000, Israel}

\date{\today}

\begin{abstract}
In this paper we present a comprehensive study of the step-emulsification process for high-throughput production of (sub-)$\mu$m-size monodisperse droplets. The microfluidic device combines a Hele-Shaw nanofluidic cell with a step-like outlet to a deep and wide reservoir. The proposed theory based on Hele-Shaw hydrodynamics provides the quasi-static shape of the free boundary between the disperse liquid phase engulfed by the co-flowing continuous phase prior to transition to oscillatory step-emulsification at low enough capillary number, Ca. At the transition the proposed theory anticipates a simple condition for critical Ca as a function of the Hele-Shaw cell geometry. The transition threshold is in excellent agreement with experimental data. A simple closed-form expression for the size of the droplets generated in step-emulsification regime derived using simple geometric arguments also shows a very good agreement with the experimental results.

\end{abstract}


\maketitle

\noindent\emph{Introduction.}
Droplet-based or digital microfluidics is a fast growing interdisciplinary research area \cite{review1,review2}. Many droplet-based microfluidic applications and technologies require high-throughput  generation of monodisperse micro-droplets of controllable size. The idea to exploit the transition from confined to unconfined flow for micro-droplet generation, known as step-emulsification (SE), was first introduced in \cite{nakajima97}. A narrow rectangular inlet channel leads to a wide and deep reservoir. The dispersed phase (non-wetting the channel walls) expands to form a tongue which grows until it reaches the step-like formation at the entrance to the  reservoir. At the step the tongue expands into unconfined spherical droplet that pinches-off from the tongue. In the past years various modifications of the step-emulsification technology were studied by this group, including  parallelization of inlet channels to form emulsification membrane for high-throughput droplet production \cite{nakajima02,nakajima06}. Other developments include droplet generation driven by a smooth ``confinement gradient", i.e. gradually varying depth in \cite{danglar13}, as opposed to a sudden (step-like) change and introduction of the co-flowing continuous phase \cite{PHS06,mall10,SBE11} capable of generation of highly monosidperse $\mu$m- and sub-$\mu$m-size drops (down to attoliter droplets).

Despite the substantial technological progress, the theoretical description of SE is very limited. In \cite{SNIS01} is was recognized that the reduction in the Laplace pressure in the tongue when the droplet expands beyond the step is responsible for the pinch-off. This mechanism was described by considering the free energy of the system. The ability of a droplet to spontaneously pinch-off/detach was calculated from the reduction in total interfacial area from before and after the droplet forms through estimating them from video images obtained using the setup described in \cite{nakajima97}. Similar arguments were put forward in \cite{evolver04}, where the finite element software (Surface Evolver) was used to identify the point of droplet pinch-off (as the minimum of interfacial energy) and predict the droplet size in membrane emulsification process under the assumption of quasi-static evolution.

More rigorous mathematical formulation relying on similar arguments is given in \cite{danglar13}, whereas the explicit expression for the capillary ``confinement gradient" force was derived for gently varying depth of the shallow inlet channel. It was found that, in contrast to classical gravity dripping problem (e.g. \cite{grav1,grav2}), the surface-tension driven droplet formation is \emph{purely geometric}, i.e. the droplet size is independent of surface tension (provided that viscous, inertial, gravity, pressure gradients and other forces are negligibly small with respect to the surface tension forces).

In this work we shall focus on SE microfluidic technology facilitated by the co-flowing continuous phase, as was first proposed by Priest et al. \cite{PHS06} where an abrupt (step-like) variation in depth of the shallow Hele-Shaw (HS) channel was exploited to produce droplets as small as $\sim$80~$\mu$~m in diameter. No extra step-like structure of \cite{nakajima97} is required in such case, as the confined quasi-2D tongue/stream of the inner liquid, completely engulfed by the co-flowing continuous phase is generated upstream within the HS cell at the T-junction and it extends towards the entrance to a deeper channel (i.e. ``the step"). The upstream width of the engulfed tongue is controlled by the flow rate ratio of two phases and their viscosities (see the detailed analysis below). Increasing the flow rate ratio (disperse-to-continuous) they observed three distinct regimes for droplet production: (i) at low values of this ratio the breakup of the confined stream inside the shallow channel occurred right after the disperse phase is injected at the T-junction; (ii) at large values of this ratio the inner (organic) phase adopts a quasi-steady tongue shape in the Hele-Shaw cell feeding a large droplet (or the balloon of the size considerably exceeding the depth of the Hele-Shaw channel) beyond the step through a short unconfined jet (the so-called ``jet-emulsification" regime). The unconfined jet is slowly thinning as the balloon grows and eventually breaks up (probably by the mechanism similar to that in \cite{LS98});  (iii) at intermediate flow rate ratio the fast dripping occurs at the step resulting in high-throughput production of highly monodisperse microdrops (named as ``step-emulsification" regime) of the size comparable to the depth of the Hele-Shaw channel. The transition between the regime (i) and (ii) was identified with the flow rate ratio corresponding to a critical width of the inner (organic) stream equal to the depth of the channel, so that diminishing the flow rate ratio the confined quasi-2D tongue becomes a cylindrical thread undergoing fast breakup due to Rayleigh-Plateau-type capillary instability.

However, the two key features of the SE process in \cite{PHS06} remained unanswered: (i) the mechanism/description of the capillary-number-dependent transition between SE and the ``jetting" regimes; (ii) surprising controllability of the SE process: increasing the flow rates of the two fluids proportionally (i.e. increasing the capillary number) results in the linear increase in the droplet generation frequency, while the size of the produced droplets remains unaltered.

A nanofluidic SE device, capable of high-throughput production of highly monosidperse (sub-)$\mu$m size (femtoliter) drops, was later proposed by \cite{mall10}. It was suggested that \emph{capillary focusing} of the engulfed tongue tip at the step controls the size of the drops. The simplified analysis of the shape of the confined quasi-2D tongue, governed by a combination of viscous and capillary forces, showed that decreasing of the capillary number yields narrowing of the quasi-static tongue's tip in accord with the experiment.

In this paper we provide a comprehensive study of nanofluidic step-emulsification facilitated by co-flowing phases, combining theory and experiment, explaining the balloon-SE transition and offering a simple quantitative prediction for the droplet size as a function of operating parameters.

\begin{figure}[tb]
\begin{center}
\includegraphics[scale=1]{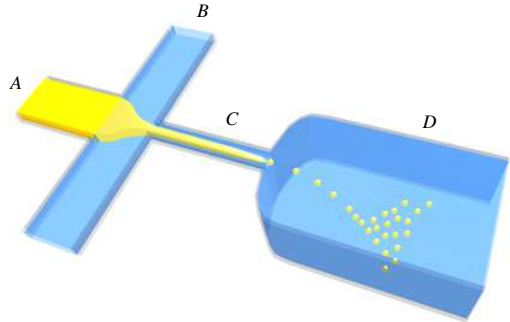}
\caption{\label{fig:schematic} Schematics of the step-emulsification nanofluidic device:  inlet of disperse phase ($A$), inlets of the aqueous phase ($B$), nanofluidic Hele-Shaw channel ($C$), collecting reservoir ($D$).  }
\end{center}
\end{figure}

\noindent\emph{Experimental setup.}
The device introduced in \cite{mall10} and used in this work is schematically shown in Fig.~\ref{fig:schematic}. Two immiscible liquids are driven through three inlets (denoted A and B in Fig.~\ref{fig:schematic}) towards a cross-junction and co-flow into a straight Hele-Shaw nanofluidic channel $C$. The inner (organic) stream does not wet the walls of the channel and thus there is a lubricating film of the continuous (aqueous) phase at the walls for all times. The two co-flowing streams arrive at a deep and wide reservoir $D$ (as compared to a deeper channel of about the same width as the HS channel in \cite{PHS06}).
At low enough flow rates SE regime takes place, where the tongue's tip undergoes oscillatory dripping at the step (entrance to the reservoir), generating droplets of nearly identical size. The droplet production frequency varies between tens to several kHz. In contrast to multistep splitting \cite{link04} such device allows generating microdrops of sub-$\mu$m size with very low polydispersity (less than 1\% by volume) in a single step \cite{mall10}. Similarly to \cite{PHS06}, increasing the flow rates of both phases results, at some point, in transition to the balloon regime.

The geometry of the  of the device used in the present study is as follows: the height/depth of the HS cells is $b=4.3$~$\mu$m, while the width was $w=38$~$\mu$m and $w=140$~$\mu$m (aspect ratio $w/b=8.8$ and $32.6$, respectively). The reservoir's height was 200~$\mu$m and its width 1.5~mm. The standard PDMS micro fabrication process was used. The motifs of the channels were formed on a silicon wafer by a photolithographic technology, a layer of PDMS (Sylgard) was cross-linked on the wafer, so that the motifs were printed onto PDMS. A thin layer of PDMS was spin-coated on a glass slide. PDMS with channel motif and the PDMS slide were combined by plasma treatment, which also rendered the surface hydrophilic. The channel is to be subjected to experiments immediately upon fabrication. Each inlet of the channel was connected to a syringe (SGE Analytical Science 100~$\mu$l). The flow rates of the fluids are controlled by a high-precision syringe pump (Nemesys) with minimum controlled flow rate of 1.32~nl/min.
The disperse/organic phase is fluorinated oil (3M 'Fluorinert' Electronic Liquid FC3283) with dynamic viscosity of $\sim 1.4$~mPa$\cdot$s, and the continuous/aqueous phase is the 23.4 g/l solution of Sodium Dodecyl Sulfate (Sigma Aldrich) in deionized water with viscosity of $\sim 1$~mPa$\cdot$s. The interfacial tension of $\gamma \approx $ 17.86~mN/m was measured by surface tension measuring instrument (KRUSS), using the ``pendent drop" method (oil droplet immersed in the aqueous phase). The formation of droplets was recorded with a fast camera (Photron Fastcam SA3) through a Zeiss microscope.

\begin{figure}[tb]
\begin{center}
\includegraphics[scale=0.9]{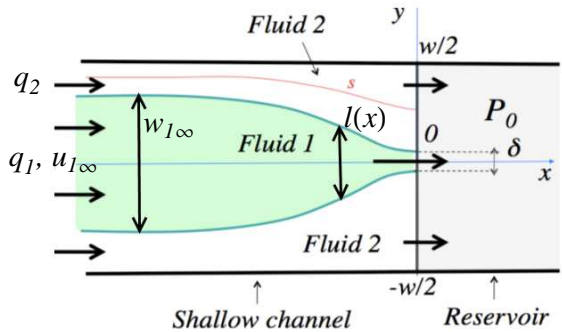}
\caption{\label{fig:HS1} Quasi-steady shape of the tongue (in the balloon emulsification regime) in a Hele-Shaw cell.}
\end{center}
\end{figure}

\noindent\emph{Problem formulation.}
Let us start with description of the quasi-steady shape of the tongue in the balloon regime. The tongue's shape is shown schematically in Figure~\ref{fig:HS1}. The depth-averaged velocity fields in two immiscible liquids (1- inner organic phase, 2 - outer aqueous phase) are governed by the depth-averaged 2D Hele-Shaw equations
\begin{equation} \label{eq:HSgeneral}
{\bm v}_{i}=-\frac{b^{2}}{12\mu_{i}}\nabla p_{i},
\end{equation}
in which $p_{i}(x, y)$ is the pressure in the region occupied by the fluid $i$ of viscosity $\mu_{i}$, and $b$ is the cell depth. Here we assume that changes in the flow direction are gentle (the validity of the assumption will be discussed later based on relevant scaling) and thus the pressure across the channel width is constant (as compared to the Saffman-Taylor ``finger" \cite{ST58}). Thus, we can express the flow rates in both phases as (with no summation on $i$):
\begin{equation} \label{eq:flrategeneral}
q_{i}=-\frac{b^{3}w_{i}}{12\mu_{i}}\frac{dp_{i}}{dx},\;\;\; i=1,2,
\end{equation}
where $w_{i}(x)$ is the local cell width occupied by the $i$th liquid.

If we define the inner fluid width as $w_{1}\equiv l(x)$  and the outer as $w_{2}\equiv w-l(x)$, with the channel width $w$, the flow rates of the two liquids (kept fixed in the experiments) read
\begin{equation} \label{eq:flrate}
q_{1}=-\frac{b^{3} l}{12\mu_{1}}\frac{dp_{1}}{dx},\;\;\; q_{2}=-\frac{b^{3}(w-l)}{12\mu_{2}}\frac{dp_{2}}{dx}.
\end{equation}

At the interface between the two phases the pressure difference is equal to the capillary pressure, that is a combination of the approximately constant Laplace pressure due to a transverse (off-plane) menisci (recall that the organic inner phase does not wet the cell walls) and the pressure due to in-plane interfacial curvature varying with $x$,
\begin{equation} \label{eq:Laplace1}
p_{1}-p_{2} =\frac{2\gamma}{b}- \frac{\gamma\:l_{xx}}{(1+l_x^2)^{3/2}} \approx \frac{2\gamma}{b}-\gamma l_{xx}\:,
\end{equation}
where the last approximate equality holds assuming gentle variation of the interface shape, $l_x^2\ll 1$. Differentiating the last equation with respect to $x$, and substituting the pressure drop evaluated from \eqref{eq:flrate} for the inner and outer liquids, we arrive at the following nonlinear ODE for the tongue's shape:
\begin{equation} \label{eq:Laplace2}
\gamma l_{xxx}=
\frac{12\mu_{1}q_{1}}{b^{3}l}-\frac{12\mu_{2}q_{2}}{b^{3}(w-l)}\;.
\end{equation}
Upstream from the step, at $x\rightarrow-\infty$ we have $l_{xxx} \rightarrow 0$ and $l\rightarrow {w_1}_\infty$, so from \eqref{eq:Laplace2} we can readily find the exact solution for the parallel co-flowing streams in a HS cell \cite{stone09}:
\begin{equation} \label{eq:Laplace3}
\frac{{w_1}_\infty}{w}=\frac{1}{1+k},\;\;\;k\equiv \frac{\mu_{2}q_{2}}{\mu_{1}q_{1}}.
\end{equation}
Introducing dimensionless variables spatial $\eta=l/w$, $\xi=x/w$, Eq.~\eqref{eq:Laplace2} can be written as
\begin{equation} \label{eq:Laplace4}
\epsilon^{-1}\:\eta_{\xi\xi\xi}=\frac{1}{\eta}-\frac{k}{1-\eta},
\end{equation}
where the \emph{modified capillary number} $\epsilon$ is defined via
\begin{equation} \label{eq:Laplace5}
\epsilon=
\frac{12\mu_{1}q_{1}}{\gamma bw}\left(\frac{w}{b}\right)^{2}\equiv \mathrm{Ca}\left(\frac{w}{b}\right)^{2}.
\end{equation}
Note that the regular capillary number, $\mathcal{C}={u_1}_\infty \mu_1/\gamma$, defined with the mean upstream velocity of the inner phase ${u_1}_\infty$ and used in \cite{PHS06}, is related to Ca as $\mathrm{Ca}=12 \mathcal{C}/(1+k)$. Note also that the flow is governed by $\epsilon$, which is equal to Ca multiplied by a large parameter $(w/b)^2 \gg 1$, emphasizing the importance of the viscous forces (due to large transverse velocity gradients) in the confined geometry. In other words, in s HS cell the flow dominated by the surface tension requires not just $\mathrm{Ca}\ll 1$, but a more restrictive condition $\mathrm{Ca} \ll (b/w)^2\ll 1$ \cite{note1}.
It can be readily seen from (\ref{eq:Laplace4}) that $l_x=\mathcal{O}(\epsilon^{1/3})$ and, therefore, the assumption of gentle variations in the flow direction requires $\epsilon^{1/3}< 1$, similarly to the well-known thin film lubrication equation (e.g. \cite{bretherton61}). Analogous approach was applied to derive the nonlinear time-dependent ODE governing thinning of the confined symmetrical neck in the HS cell in \cite{CDGKSZ93}. However, in most practical cases \cite{PHS06,mall10} $\epsilon$ is not small (even if Ca is small) due to the large factor $(w/b)^2$ multiplying Ca. One may consider using the full expression for the interfacial curvature in Eq.~\ref{eq:Laplace1} to approximate the solution when the underlying assumption of nearly unidirectional flow (i.e. $\epsilon <1$) is violated, yielding
\be
\label{eq:Laplace4a}
\epsilon^{-1}\:\partial_\xi\left(\frac{\eta_{\xi\xi}}{(1+\eta_\xi^2)^{3/2}}\right)=\frac{1}{\eta}-\frac{k}{1-\eta}\:.
\ee
The parallel flow solution (\ref{eq:Laplace3}) of Eq.~\ref{eq:Laplace4} (or Eq.~\ref{eq:Laplace4a}) then reads $\eta_\infty={w_1}_\infty/w=1/(1+k)$ as $\xi\rightarrow-\infty$.

At some distance upstream from the step, the width of the tongue $\eta$ starts to deviate from constant $\eta_\infty$, so we can write $\eta=\eta_\infty+\widetilde{\eta}$, where $\widetilde{\eta} \ll \eta_\infty$ is a small perturbation. When $\eta$ is close to $\eta_\infty$ the Eq.~\ref{eq:Laplace4} (or Eq.~\ref{eq:Laplace4a}) can be linearized to read
\begin{equation} \label{eq:Laplace8}
\epsilon^{-1}\: \widetilde{\eta}_{\xi\xi\xi}  +\frac{(1+k)^3}{k} \widetilde{\eta}=0\:,
\end{equation}
Eq.~\ref{eq:Laplace8} can be readily solved to give
\be \label{eq:asymp1}
\eta \approx \frac{1}{1+k}+\beta \re^{\frac{1}{2}\lambda \xi} \cos{\left(\frac{\sqrt{3}}{2}\lambda \xi\right)}\:,
\ee
where $\lambda=(1+k) \left(\frac{\epsilon}{k}\right)^{1/3}$ and $\beta$ is an integration constant. In deriving (\ref{eq:asymp1}) we used the requirement  $\eta\rightarrow 1/(1+k)$ as $\xi \rightarrow -\infty$, and the invariance of the solution to the shift in the origin, so there is one free constant $\beta$. This constant is determined by the outlet conditions, i.e. $\eta_\xi=0$ and prescribed curvature $\eta_{\xi\xi}$ at some \emph{a priori} unknown axial distance $\xi_*$. If we follow \cite{mall10} and assume that at the entrance to the reservoir the pressures in both phases equilibrate, and since the pressure jump across between the interface at the entrance to the reservoir is determined by Eq.~\ref{eq:Laplace1} with $p_1-p_2=0$, we arrive at $\eta_{\xi\xi}=2(w/b)$ at the step at $\xi_*$. However, the pressure in the unconfined thread beyond the step feeding the balloon is somewhat higher than the ambient pressure of the continuous phase in the reservoir. However, a more general condition $\eta_{\xi\xi}=2c(w/b)$ at $\xi_*$ is expected, where the dimensionless parameter $c\le 1$ is controlled by dynamics of capillary breakup of unconfined viscous thread surrounded by another viscous liquid and connecting the confined tongue and the unconfined balloon \cite{LS98}. \\

\begin{figure}[tb]
\begin{center}
\begin{tabular}{cc}
\includegraphics[scale=0.78]{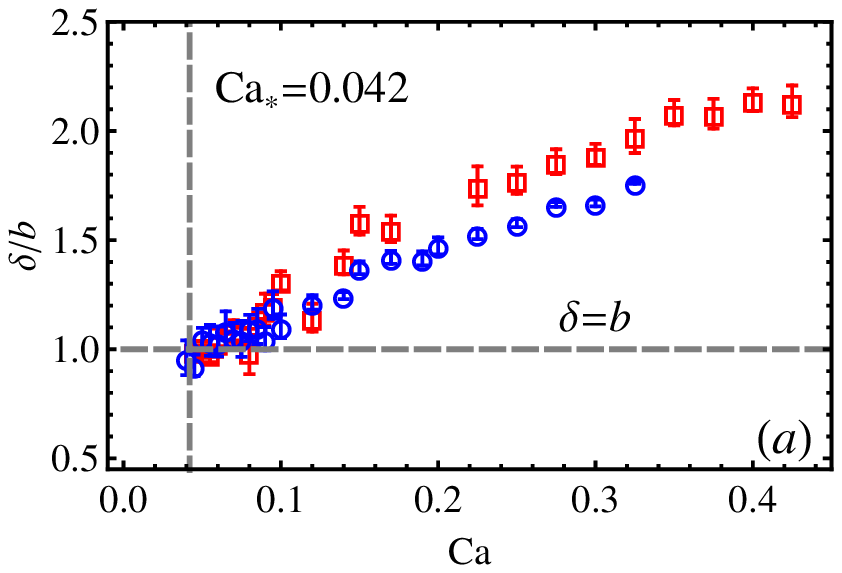}\\
\;\;\;\;\;\includegraphics[scale=0.8]{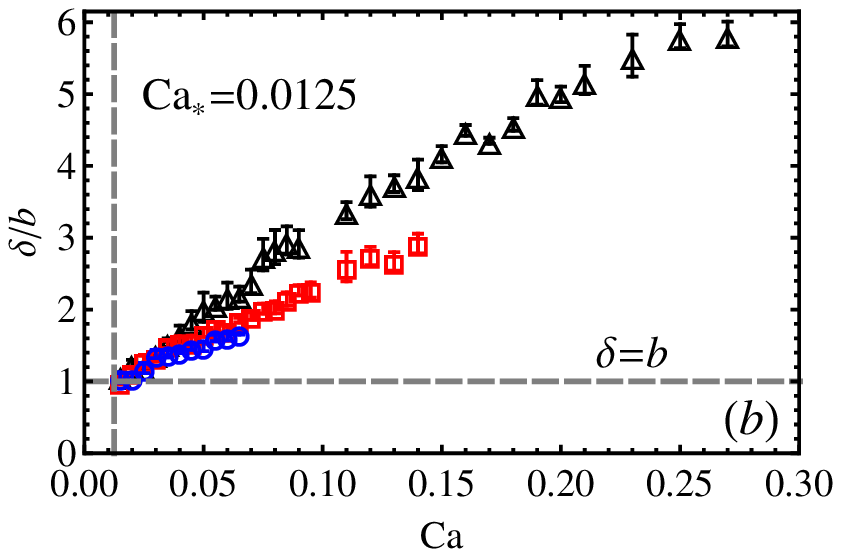}
\end{tabular}
\caption{\label{fig:width} The scaled tongue width at the outlet $\delta/w$ vs. Ca in a balloon regime in a HS cell (a) $w/b=8.8$ and $k=0.5$ ($\square$), $k=1$ ($\circ$); (b) $w/b=32.6$ and $k=1$ ($\vartriangle$), $k=3$ ($\square$), $k=6.5$ ($\circ$); the dashed lines denote the transition threshold to step-emulsification regime corresponding to $\delta\simeq b$.}
\end{center}
\end{figure}

\noindent\emph{Balloon-SE transition and phase diagram.}
Our experimental findinds confirm that near at transition threshold the critical tongue's width is about the height of the HS cell, $\delta\sim b$. Figs.~\ref{fig:width}\emph{a}, \emph{b} show the scaled width of the tongue $\delta/b$ in the quasi-static balloon regime vs. $\mathrm{Ca}=\frac{12\mu_{1}q_{1}}{\gamma bw}$ in two HS cells with the aspect ratio $w/b=8.8$ and $32.6$, respectively, for different values of $k$. It can be readily seen that in each cell, the transition to dripping occurs at the critical $\mathrm{Ca}_*=0.042$ and $0.0125$, respectively, for all $k$'s so that $\delta/w\simeq 1$. Increasing Ca above $\mathrm{Ca}_*$ yields steadily growing width of the quasi-2D tongue, with rate of the grows depending on $k$. These results indicate that the capillary instability of the cylindrical jet as $\delta \sim b$ is likely to be responsible for the transition from the quasi-steady balloon regime to oscillatory SE regime. Similar mechanism was suggested to be operative in breakup of the narrow tongue, i.e. ${w_1}_\infty\sim b$, upstream well inside the HS cell at low Ca's in \cite{PHS06}.

\begin{figure}[t]
\begin{center}
\begin{tabular}{cc}
\includegraphics[scale=0.52]{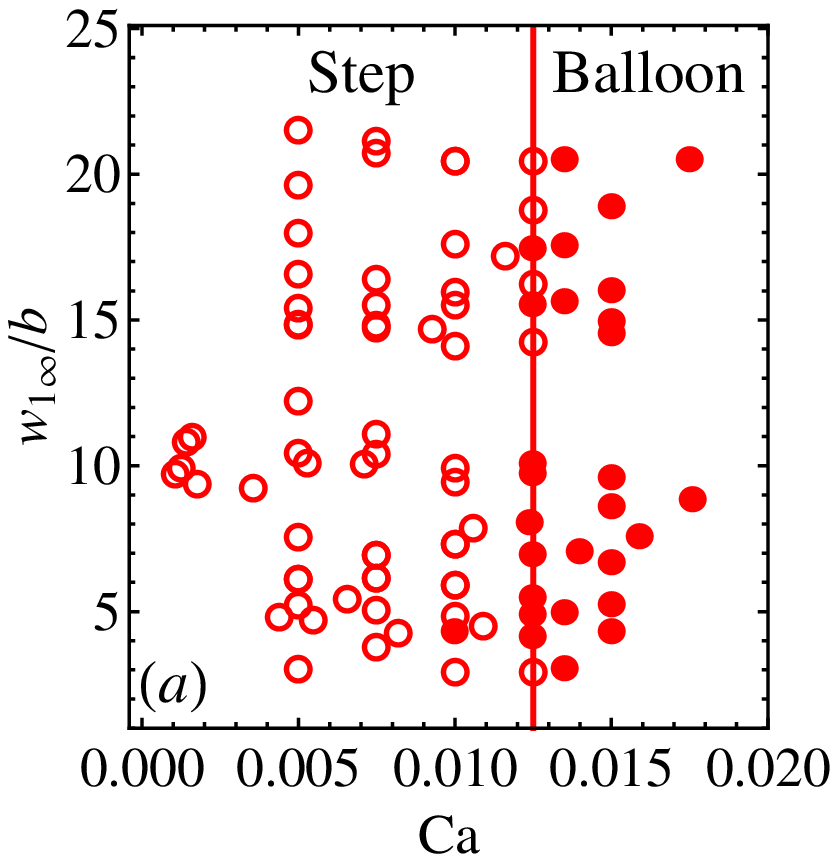} &
\includegraphics[scale=0.52]{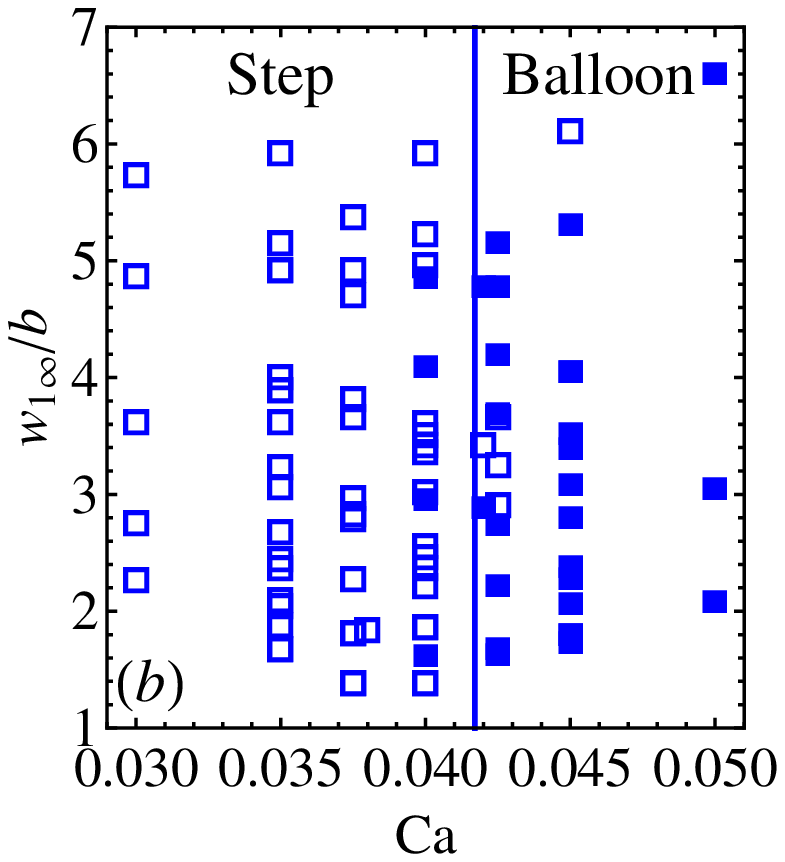} \\ \\
\end{tabular}
\includegraphics[scale=0.85]{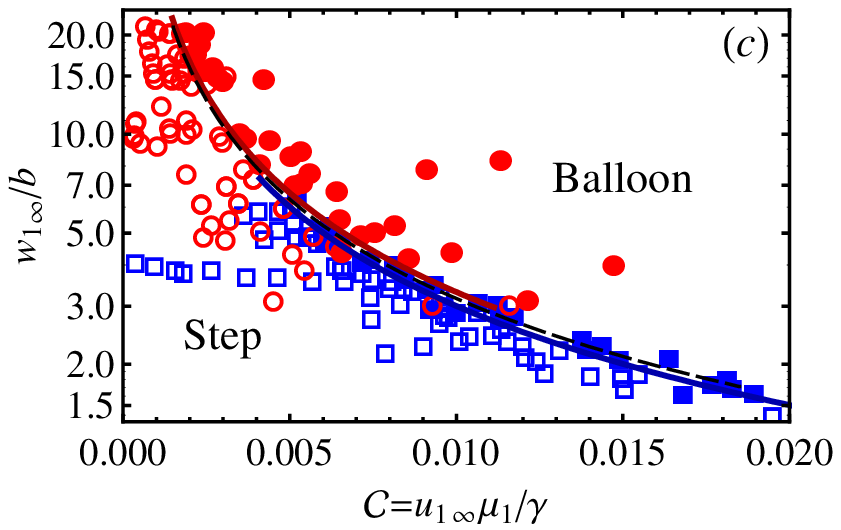}
\caption{Phase diagrams; the void symbols correspond to the SE regime and the filled symbols to the balloon regime; squares and circles stands for the channel aspect ratios $w/b=8.8$ and $w/b=32.6$, respectively; (a) in terms of Ca for $w/b\simeq 32.6$, the vertical line stands for transition threshold $\mathrm{Ca}_* \simeq 0.0125$;(b) in terms of Ca for $w/b\simeq 8.8$, the vertical line stands for transition threshold $\mathrm{Ca}_*\simeq 0.042$; (c) in terms of the scaled upstream width tongue width ${w_1}_\infty/w$ vs. the regular capillary number $\mathcal{C}$ (log-linear plot); the solid (red and blue) lines are the boundaries re-drawn from Figs.~(\emph{a}) and (\emph{b}), the dashed (black) line is the transition threshold (\ref{eq:trans}) with $A=0.032$.
\label{fig:phase-diag}}
\end{center}
\end{figure}

The above experiments indicate that the transition occurs at some critical $\mathrm{Ca}_*$ that only varies the cell aspect ratio $w/b$ and independent of $k$, i.e. of the upstream width of the tongue (see Figs.~\ref{fig:phase-diag}\emph{a},\emph{b}). When the stability diagram is re-plotted in terms of $\frac{{w_1}_\infty}{b}=\left(\frac{w}{b}\right)\frac{1}{k+1}$ vs. the standard capillary number, $\mathcal{C}={u_1}_\infty\mu_1/\gamma=\mathrm{Ca} (1+k)/12$, it appears to be independent of the cell aspect ratio, $w/b$, as both boundaries in Figs.~\ref{fig:phase-diag}\emph{a},\emph{b} collapse into a single curve in Fig.~\ref{fig:phase-diag}\emph{c}. The phase diagram in Fig.~\ref{fig:phase-diag}\emph{c} also agrees well with the earlier results of \cite{PHS06}. The generic nature of the transition curve in Fig.~\ref{fig:phase-diag}\emph{a} can be realized as follows. Given the relation between the critical Ca$_*$ and $\mathcal{C}_*$, the transition threshold is given by
\be
\frac{{w_1}_\infty}{b}= \frac{A}{\mathcal{C}_*}\:,\label{eq:trans}
\ee
where $A=\frac{1}{12}\Ca_*(w/b)$, with $\Ca_*$ being the constant critical capillary that only varies with the aspect ratio. For the results (\ref{eq:trans}) to be universal, (i.e. independent of the aspect ratio), we expect that $\Ca_* (w/b)=\mathrm{Const}$. Indeed, for $w/b=32.6$ the critical capillary is $\Ca_*\approx0.012$, while for $w/b=8.8$ it was found that $\Ca_*\approx0.042$, making the product $\Ca_*(w/b)$ equal to $0.39$ and $0.37$, respectively. Therefore the SE-balloon universal transition boundary is described by
\be
\Ca_*\left(\frac{w}{b}\right )\simeq 0.38\;,\label{eq:trans2}
\ee
or alternatively by (\ref{eq:trans}) with $A\approx 0.032$.

Actually, (\ref{eq:trans2}) readily follows from scaling of the derived ODE (\ref{eq:Laplace4}) (or Eq.~\ref{eq:Laplace4a}) governing the quasi-steady tongue shape in the balloon regime. Near the transition the second term in the RHS is small with respect to the first one when $\eta \sim b/w \ll 1$ (for $w/b \gg k$). At the HS cell outlet at $\xi=\xi_*$ we have $\eta_{\xi\xi}=2\textrm{c}(w/b)$, where $\mathrm{c}$ is a dimensionless constant. To eliminate $\epsilon$ and $w/b$ from the equation and the boundary condition at the cell outlet we re-scale the tongue width and the axial distance as $\eta=\widehat{\eta} \Ca^2 (w/b)$ and $\xi=\widehat{\xi} \Ca$ to yield $\widehat{\eta}\widehat{\eta}_{\widehat{\xi}\widehat{\xi}\widehat{\xi}}=1$ and $\widehat{\eta}_{\widehat{\xi}\widehat{\xi}}=2\mathrm{c}$ at $\widehat{\xi}_*$. At the transition threshold the tongue width approaching the HS cell depth, $\eta\approx (b/w)\sim\Ca_*^2(w/b)$, or just $\Ca_* (w/b)=\mathrm{Const}$, where the value of the Const is unique provided that $c$ at the transition threshold is not varying with $k$. This is in agreement with (\ref{eq:trans2}).

\noindent\emph{Quasi-static tongue shapes.}
Besides predicting the correct transition threshold, the above theoretical model can also be used to compute the quasi-static shapes of the tongue in the balloon regime, treating $c$ as fitting parameter. We use the asymptotic upstream solution (\ref{eq:asymp1}) to derive a consistent set of initial conditions ($\eta$, $\eta_\xi$ and $\eta_{\xi\xi}$, all depending on $\beta$) for the numerical integration of Eq.~\ref{eq:Laplace4a} as initial value problem. Thus, (\ref{eq:Laplace4a}) is integrated  in the direction of increasing $\xi$ up to some (\emph{a priori} unknown) position $\xi_*$ by ``shooting" method, i.e. fitting the value of $\beta$ so that $\eta_{\xi\xi}=2c(w/b)$ at $\xi=\xi_*$ whereas $\eta_\xi=0$. The resulting value of $\eta(\xi_*)$ yields the width of the tongue at the nano-fluidic HS cell outlet in the quasi-steady balloon regime. The value of $c$ giving the critical width $\eta=b/w$ at $\xi_*$ determines the critical (negative) in-plane curvature of the tongue at the balloon-SE transition.

\begin{figure}[tb]
\begin{center}
\begin{tabular}{cc}
\includegraphics[scale=0.9]{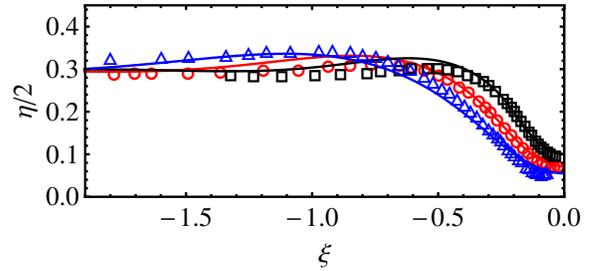}
\end{tabular}
\caption{\label{fig:shapes} Quasi-static tongue shapes, comparison between the theory and experiment for $k=0.67$ and $w/b=8.8$ for three different values of the capillary number: $\Ca=0.25$ ($\square$), $\Ca=0.1$ ($\circ$) and $\Ca=0.045(\approx \Ca_*)$ ($\vartriangle$). Solid lines stand for the computed profiles upon fitting the in-plane curvature of the tongue at the outlet at $\xi_*=0$; for $\Ca=0.045$ the value of $\eta_{\xi\xi}(0)=5.75$ (corresponding to $c\simeq 0.33$) yields outlet width $\eta(0)\simeq b/w=0.11$.
}
\end{center}
\end{figure}

The comparison between the computed profiles and the experimental data is shown in Fig.~\ref{fig:shapes} for $k=0.67$, cell aspect ratio $w/b=8.8$ and several values of capillary number: $\Ca=0.25$ ($\square$), $\Ca=0.1$ ($\circ$) and $\Ca=0.045(\approx \Ca_*)$ ($\vartriangle$). As the pressure in the thread beyond the step is unknown, the in-plane curvature at the outlet at $\xi_*=0$ was fitted (i.e. by fitting the parameter $c$) to match the experimentally observed outlet width $\eta(0)$. The value of $c\simeq 0.33$ was found to fit the critical width $\eta(0)=b/w\approx 0.11$ at $\Ca=0.045$, while for $\Ca=0.1$ and $0.25$ the corresponding best-fitted values of $c$ are $0.45$ and $0.62$, respectively.
While full 3D numerical simulations (e.g. using VoF method, \cite{alr11,LAJT12}) are required to determine in-plane curvature at the outlet in a self-consistent fashion, fitting the value of $c$ numerically to best-fit the experimentally determined outlet width yields excellent quantitative agreement between the experimentally measured and numerically computed profiles in Fig~\ref{fig:shapes}. \textit{Thus, in accord with \cite{mall10}, decreasing $\Ca$ yields better ``capillary self-focusing", i.e. narrower outlet width of the confined tongue, $\delta$, while the transition to oscillatory SE regime occurs at critical $\mathrm{Ca}_*$ at which $\delta \approx b$.}

Computing the critical profiles at $\Ca_*(w/b)=0.38$ corresponding to $\eta(\xi_*)=b/w$ (e.g. see $\vartriangle$'s in Fig.~\ref{fig:shapes}) for the HS cell with aspect ratio $w/b=32.6$, we found that for $k$ varying in a wide range $k=0.5\div5$ the critical in-plane curvature varies only slightly in a narrow interval $\eta_{\xi\xi}\approx 21.6\pm 2.3$, corresponding to $c\approx 0.33 \pm 0.03$ . This is in agreement with our previous scaling arguments in favor of the transition threshold (\ref{eq:trans2}).

\begin{figure*}[tb]
\begin{center}
\includegraphics[scale=0.55]{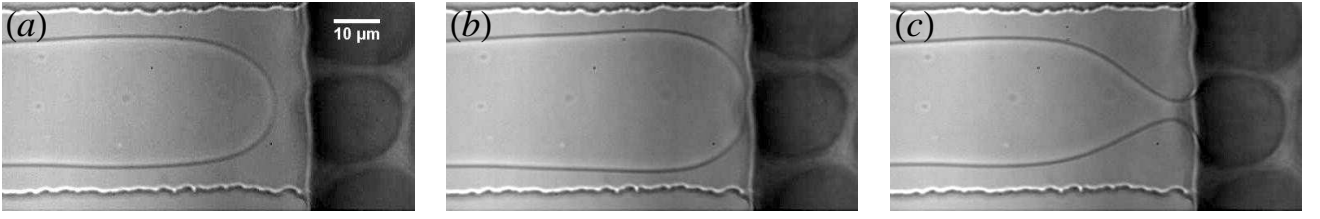}
\caption{The step-emulsification breakup regime observed in experiments using fluorinated oil (inner phase) in water (outer phase); the time progresses from left to right. The scale bar is the same for all figures. (a) the tongue right after the droplet pinch-off and retreat; (b) the tongue just reached the step; (c) the tongue shape right before pinch-off and retreat; \label{fig:step}}
\end{center}
\end{figure*}

\noindent\emph{Droplet size prediction in SE regime.}
The periodic evolution process of the droplet generation in step-emulsification regime is shown in Figs.~\ref{fig:step}. First, the tongue flows under the applied pressure gradient in the HS cell towards the step (Fig.~\ref{fig:step}\emph{a}, \emph{b}). The time $\Delta t_{a\rightarrow b}$ it takes for the tongue to reach the step is controlled by the flow rates of both phases. After the tongue reaches the step (as in Fig.~\ref{fig:step}\emph{b}) the inner fluid is pushed into the reservoir forming a droplet, the droplet formation is accompanied by the fast narrowing of tongue's tip and considerable transverse velocities (see Fig.~\ref{fig:step}~\emph{b},\emph{c}). The previous assumptions of nearly unidirectional flow in this case are obviously violated and the developed quasi-steady theory does not apply in SE regime. We denote by the \emph{dripping time} the time is takes for the tongue at the step to deform up to the pinch-off, i.e. $\Delta t_{b\rightarrow c}\equiv\Delta t_\mathrm{drip}$. As we discussed in the introduction the driving force behind the step-emulsification process is the surface tension or \emph{confinement gradient} \cite{danglar13} whereas the formation of the unconfined 3D droplet at the step reduces the total interfacial area/energy. Droplet formation is accompanied by the reduction of the inner pressure $p_1$ and thinning/squeezing of the neck feeding the droplet by a higher outer pressure $p_2$. The neck in Fig.~~\ref{fig:step}\emph{c} eventually pinches-off (presumably due to a capillary instability), followed by a very fast retreat/recoil of the tongue within the HS cell as shown in Fig.~\ref{fig:step}\emph{a}.

The volume of the droplet formed in between Figs.~\ref{fig:step}b,c has two contributions: one from the deformation of the tongue, pushing some liquid out during the time between 'b' and 'c', and the second is due to constant inner liquid influx, equal to the dripping time, $\Delta t_{\mathrm{drip}}$, times  $q_1$:
\be
V_d=\Delta V_{\mathrm{def}}+\Delta t_{\mathrm{drip}}\: q_1. \label{eq:step}
\ee
The first term due to tip narrowing can be estimated from mass conservation as a difference between tongues' volumes in Figs.~\ref{fig:step}b and \ref{fig:step}a (i.e. volume increment),$V_{\mathrm{def}}= V_b-V_a$, assuming very fast retreat so that the volume of the tongue in 'c' and 'a' is the same.

\begin{figure}[tb]
\begin{center}
\begin{tabular}{cc}
\includegraphics[scale=0.8]{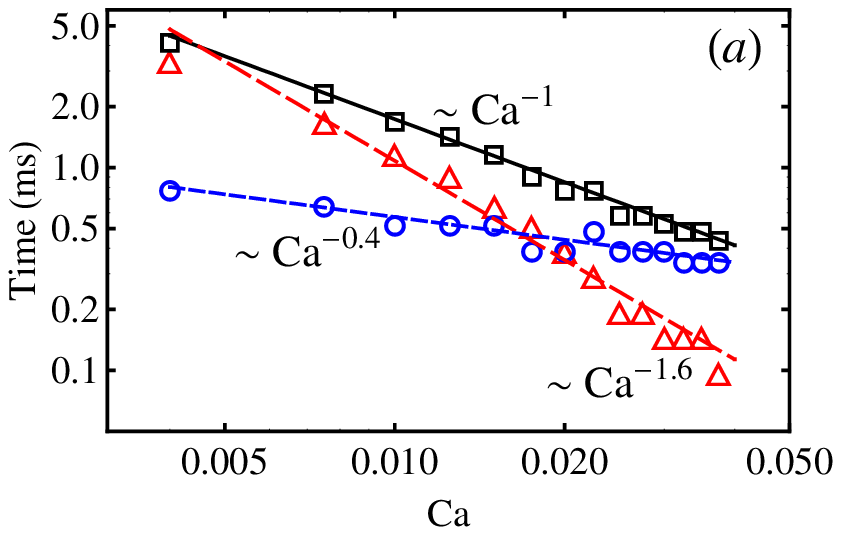}\\
\includegraphics[scale=0.8]{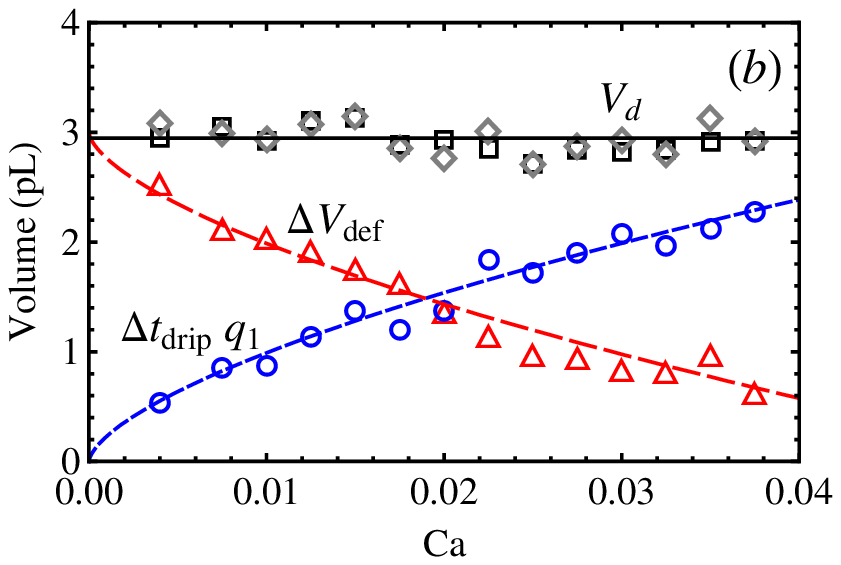}
\end{tabular}
\caption{Characteristics of step-emulsification regime in HS cells with $w/b=8.8$ for $k=2$: (a) the droplet formation times (ms) vs. Ca (log-log plot): $\Delta t_{a\rightarrow b}$ ($\vartriangle$), $\Delta t_\mathrm{drip}$ ($\circ$); full time period $T=1/f$ ($\square$). The lines are the best power-law trends. (\emph{b}) Individual volumes (picoliter) vs. Ca: deformation volume $\Delta V_\mathrm{def}$ ($\vartriangle$), influx volume $\Delta t_\mathrm{drip} q_1$ ($\circ$), their sum $\simeq 2.98\pm 0.14$~pL ($\diamond$) and measured droplet volume $V_d=2.95\pm 0.11$~pL ($\square$). The dashed (short and long dashes) are the Eqs.~\ref{eq:volumes}, respectively, with $\alpha\simeq 6.2$ and $\nu\simeq 0.63$.
\label{fig:time}}
\end{center}
\end{figure}

Depending on the upstream width of the tongue ($k$), the relative importance of these two terms can interchange, e.g. for $k<1$ (wide tongues) the second term (due to influx) is dominant while for $k>1$ (narrow tongues) the first term (due to deformation) is dominating, can be seen in experiments. Both terms on the RHS of (\ref{eq:step}) depend on the flow conditions. It is reasonable to assume that for narrow tongues, the second term in (\ref{eq:step}) is small (the influx due to $q_1$ through a narrow tongue is small during $\Delta t_\mathrm{drip}$) and the droplet size is mainly controlled by the deformation of the tongue. In \cite{PHS06} in Fig.2b the experimental results correspond to $k\approx2.1$, while our experiments indicate that at $k\sim2$ the contribution of the influx to the volume is only about 20\% of the droplet volume. For wider tongues the contribution of the 2nd term on the RHS in (\ref{eq:step}) to the droplet volume could be considerable, e.g. our experiments suggest it is about 75\% for $k=0.5$. Analogously, for low Ca, the 1st term is expected to contribute the most to the droplet volume, while for high Ca the 2nd term should be dominant. However, our experiments show (in accord with previous works, e.g. \cite{PHS06},\cite{SBE11}) that the droplet volume is independent of Ca for a fixed flow rate ratio $q_2/q_1$ (i.e. fixed $k$). More detailed analysis of the experimental data verify that the Ca-dependent contributions to the droplet volume from both terms in the RHS of (\ref{eq:step}) cancel out, i.e.
\be
\Delta V_\mathrm{def}\sim V_d(1-\alpha \mathrm{Ca}^\nu)\,,\quad \Delta t_\mathrm{drip}\: q_1\sim \alpha V_d \mathrm{Ca}^\nu\;,\label{eq:volumes}
\ee
where $\alpha$, $\nu$ are some dimensionless constants. Our experimental results show that both $\Delta t_{a\rightarrow b}$ and $\Delta t_\mathrm{drip}$ depend on Ca, as intuitively expected for confined flow governed by interplay of viscous and capillary forces. These two times are shown in Fig.~\ref{fig:time}\emph{a} in a HS cell with $w/b=8.8$ for $k=2$ as a function of Ca ($\circ$ and $\vartriangle$). These results demonstrate that $\Delta t_\mathrm{drip}\sim Ca^{-0.4}$ while $\Delta t_{a\rightarrow b}\sim \mathrm{Ca}^{-1.6}$. Their sum gives a dripping period $T
\sim \mathrm{Ca}^{-1}$ ($\square$).

The corresponding individual experimentally measured volumes  $\Delta V_\mathrm{def}$ and $\Delta t_\mathrm{drip} q_1$ are depicted vs. Ca in Fig.~\ref{fig:time}\emph{b}. The deformation volume was estimated as $\Delta V_\mathrm{def}=\Delta S_\mathrm{def} b$, where $ \Delta S_\mathrm{def}\approx S_b-S_a$ is the corresponding surface area increment (see Figs.~\ref{fig:step}\emph{a},\emph{b}). Since $\Delta t_\mathrm{drip}\sim Ca^{-0.4}$ and  $q_1\sim \mathrm{Ca}$, the volume due to influx $\Delta t_\mathrm{drip} q_1\sim \mathrm{Ca}^{\nu}$ with $\nu\approx 0.6$ as can be readily seen (long-dashed line). The deformation volume $\Delta V_\mathrm{def}$ follows the second Eq.~(\ref{eq:volumes}) so that the total droplet volume is constant, $V_d\approx 2.9$~pL, and independent of Ca.

\begin{figure}[tb]
\begin{center}
\includegraphics[scale=0.9]{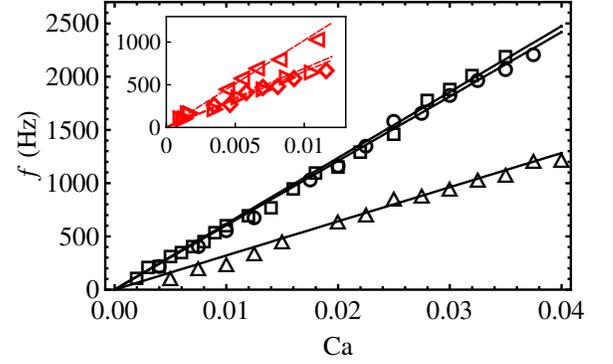}
\caption{The droplet production frequency, $f$, in SE regime vs. Ca in HS cells with $w/b=8.8$: $k=0.5$ ($\vartriangle$), $k=2$ ($\circ$), $k=2.5$ ($\square$); the inset shows the results for
$w/b=32.6$: $k=1$ ($\diamond$), $k=2.5$ ($\vartriangleright$) and $k=5.2$ ($\vartriangleleft$). The continuous lines are the best linear fits.  \label{fig:freq}}
\end{center}
\end{figure}

Since the size of the generated droplet in the SE regime is independent of Ca, obviously the frequency of their production, $f$ (i.e. number of droplets per unit time), should grow linearly with Ca (or with $q_1$ in \cite{PHS06}). Indeed, from mass conservation $f=q_1/V_d$. Re-writing $q_1$ via Ca we obtain $f=\mathcal{K} Ca$, where the proportionality constant $\mathcal{K}=\gamma b w/2\pi \mu_1 d^3$. The experimental results depicted in Fig.~\ref{fig:freq} confirm the linear growth of $f$ with Ca in agreement with previous results \cite{PHS06}. Analogously, the dripping period $T=1/f\sim \mathrm{Ca}^{-1}$ in agreement with ~Fig.~\ref{fig:time}\emph{a} ($\square$, solid line).
\emph{Thus, even though the dynamics within the HS cell is governed by the combination of the viscous and capillary forces (making the balloon-SE transition threshold Ca-dependent, see Figs.~\ref{fig:phase-diag}), this dynamics seem to be slaved to the droplet formation in the reservoir controlled entirely by the surface tension as the viscous forces are negligibly small.}

Postponing the intriguing question concerning cancelation of the Ca-dependent contributions to the droplet volume for future investigation, we shall focus on the limit  $\Ca\rightarrow 0$ where the droplet volume is controlled predominantly by tongue deformation, i.e. $V_d\approx \Delta V_{\mathrm{def}}$. In this case it is reasonable to assume that the droplet is spontaneously formed once the total interfacial energy (i.e. area) is lowered, i.e. droplet formation is favored thermodynamically. Simple arguments based on comparison of corresponding interfacial areas and volumes yield the critical droplet size. $V_{\mathrm{def}}$ can be estimated from mass conservation as a difference between tongues' volumes in Figs.~\ref{fig:step}\emph{b} and \ref{fig:step}\emph{a} (i.e. volume increment), $V_\mathrm{def}= V_b-V_a$, assuming very fast retreat so that the volumes of the tongue in 'c' and 'a' are essentially the same, $V_a\approx V_c$. In such case we need to compare the interfacial area in 'b' with that in 'c', where the latter is equal to the area of the tongue in 'a' plus the surface area of the droplet produced. Let us denote by $\ell$ the distance over which the tongue retreats backward into the HS cell after breakup, i.e. from 'c' to 'a' (or advances from 'a' to 'b'). therefore the increment in the interfacial area (``new" interface) between 'a' and 'b' (assuming that no deformation occurs, just translation of the tongue forward) reads
\[
\Delta S_{a\rightarrow b}\approx 2 {w_i}_\infty \ell+2\ell b\:,
\]
where the first term stands for the in-plane surface and the 2nd term for the area of the off-pane menisci. On the other hand, the surface area of the droplet is just $S_\mathrm{d}=\pi d^2$. Conservation of mass requires that the volume difference $\Delta V_{a\rightarrow b}\approx {w_1}_\infty \ell b$ should be equal to the volume of the produced droplet, $V_d=\pi d^3/6$. The breakup of droplet is favorable when $\Delta S_{a\rightarrow b} \gtrsim S_d$. Solving the equation $\Delta S_{a\rightarrow b}= S_d$ together with mass conservation constraint, $\Delta V_{a\rightarrow b}=V_d$ yield the diameter of the \emph{smallest} droplet and the corresponding distance $\ell$
\be
d=\frac{3 b {w_1}_\infty}{(b+{w_1}_\infty)}\:,\quad \ell=\frac{9 b^2 \pi {w_1}_\infty^2}{2 (b+{w_1}_\infty)}\:.
\ee

Using the upstream parallel-flow solution, $\frac{{w_1}_\infty}{w}=\frac{1}{1+k}$, the scaled droplet diameter $d/b$ reads
\be
\frac{d}{b}=3\left[1+\frac{(1+k)}{w/b}\right]^{-1}\:. \label{eq:ddrop}
\ee

This results indicates that the diameter $d$ of the smallest droplet that can be produced in SE regime varies between $\approx 2b\div3b$. This is in qualitative agreement with previously experimental observations reporting the smallest droplet diameter two to three times the height the inlet channel height \cite{nakajima02,mall10,SBE11}. The narrower inner streams (i.e. lower ${w_1}_\infty$ and higher $k$) or/and larger aspect ratio $w/b$ results in smaller droplets.

\begin{figure}[tb]
\begin{center}
\includegraphics[scale=0.8]{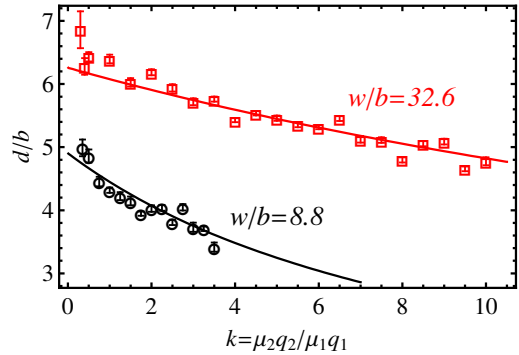}
\caption{The diameter of the droplets, $d/b$, produced by step-emulsification in the HS cells with two aspect ratios: $w/b=8.8$ (black circles) and $w/b=32.6$ (red squares). The solid lines are the prediction in Eq.~(\ref{eq:ddrop}) multiplied by $1.85$ and $2.15$ respectively.  \label{fig:drop-vol}}
\end{center}
\end{figure}

Using this solution we obtain that the interfacial area increment $\Delta S_{a\rightarrow b}\sim d^3/b$, while the interfacial area of the produced droplet is quadratic in its diameter, $S_d\sim d^2$. Therefore, production of small droplets with diameter smaller than that in Eq.~(\ref{eq:ddrop}) is unfavored since interfacial area increases. The size of the droplet produced in experiments using channels with two different aspect ratios $w/b=8.8$ and $w/b=32.6$ seem to agree well with the prediction in (\ref{eq:ddrop}) multiplied by a factor of $\approx 2$. The comparison is provided in Fig.~\ref{fig:drop-vol} showing excellent agreement between the theory and the experiment. Note that the proposed geometric construction of the tongue shape is only approximate (i.e. going from `a' to `b' the tongue does not simply advances to the right, but also inflates) and thus the appearance of the constant multiplicative factor $\approx 2$ varying slightly with the cell aspect ratio is not surprising. However, the agreement between the simple theory (\ref{eq:ddrop}) and the experimental results for a particular HS cell with given aspect ratio is excellent in a wide range of $k$. For wide tongues (e.g. $k=0.5$) the theoretical prediction (\ref{eq:ddrop}) somewhat underestimates the droplet size. Wide tongues require small capillary number $\mathcal{C}$ and thus low flow rate of the continuous phase $q_2$ in the SE regime (see the phase diagram in Fig.~\ref{fig:phase-diag}\emph{a}) and generated droplets are not conveyed fast enough further downstream in the reservoir following their pinch-off. Thus formation of the droplets is affected by their crowding near the step. The disturbance due to crowding yields relatively large variance in the measured droplet size at $k=0.5$.

\noindent\emph{Concluding remarks.} Here we provide a comprehensive experimental and theoretical study of the step-emulsification process in a microfluidic device composed of a shallow nanofluidic (Hele-Shaw) cell connected to a deep and wide reservoir. The theoretical model based on depth-averaged Hele-Shaw hydrodynamics yields the nonlinear ODE for the quasi-static shape of the confined tongue of the disperse liquid, engulfed by the co-flowing continuous phase, prior to transition to the oscillatory step-emulsification regime at low enough capillary number, Ca. At the transition threshold, the developed theory suggests a very simple condition for the critical capillary number as a function of the Hele-Shaw cell aspect ratio, $\Ca_* \; (w/b)=\mathrm{Const}$, being in excellent agreement with our experimental data showing that $\mathrm{Const}\approx 0.38$. The computed tongue shapes, determined by fitting the curvature of confined tongue of the disperse liquid at the outlet, are in excellent agreement with the experimental results. The closed-form expression for the smallest size of the droplets produced in step-emulsification regime is found using simple thermodynamic and geometric arguments as a function of flow rates, viscosities of both phases and geometry of the Hele-Shaw cell. This prediction shows an excellent agreement with our experimental findings.\\

\noindent\emph{Acknowledgement.} A.M.L. acknowledges the support of the Israel Science Foundation (ISF) via the grant \#1319/09 (``Unit operations in droplet-based microfluidics") and Joliot Chair visiting position at Ecole Sup\'erieure de Physique et Chimie de Paris (ESPCI). Z.L. and P.T. acknowledge CNRS and ESPCI for their support.


\begin{thebibliography}{99}
\bibitem{note1} A similar effect of confinement on the interplay between inertia and viscous forces concerns the condition of negligible liquid inertia, $(b/w) \mathrm{Re} \ll 1$, where the Reynolds number is defined as $\mathrm{Re}=\rho b^3 |\nabla p|/\mu^2$ \cite{batchelor67}. In other words, viscous forces dominate over inertia in confinement even at finite $\mathrm{Re}\sim \mathcal{O}(1)\ll w/b$.
\end{thebibliography}
\end{document}